\documentclass[floatfix,aps,twocolumn,preprintnumbers,superscriptaddress,floatfix,prl]{revtex4-1}
\usepackage{bm}
\usepackage{lipsum}
\usepackage{graphicx,amssymb}
\usepackage{dcolumn}
\usepackage{bm}
\usepackage{xcolor}

\begin{document}
\draft \preprint{Huang $et~al.$}

\title{Two-dimensional superconducting fluctuations associated with charge density wave stripes in La$_{1.87}$Sr$_{0.13}$Cu$_{0.99}$Fe$_{0.01}$O$_4$}

\author{H. Huang}
\affiliation{Stanford Synchrotron Radiation Lightsource, SLAC National Accelerator Laboratory, Menlo Park, California 94025, USA}
\author{S.-J. Lee}
\affiliation{Stanford Synchrotron Radiation Lightsource, SLAC National Accelerator Laboratory, Menlo Park, California 94025, USA}
\author{Y. Ikeda} 
\affiliation{Institute for Materials Research, Tohoku University, Katahira 2-1-1, Sendai, 980-8577, Japan}
\author{T. Taniguchi} 
\affiliation{Institute for Materials Research, Tohoku University, Katahira 2-1-1, Sendai, 980-8577, Japan}
\author{M. Takahama} 
\affiliation{Institute for Materials Research, Tohoku University, Katahira 2-1-1, Sendai, 980-8577, Japan}
\author{C.-C. Kao}
\affiliation{SLAC National Accelerator Laboratory, Menlo Park, California 94025, USA}
\author{M. Fujita}\email{fujita@imr.tohoku.ac.jp}
\affiliation{Institute for Materials Research, Tohoku University, Katahira 2-1-1, Sendai, 980-8577, Japan}
\author{J.-S. Lee}\email{jslee@slac.stanford.edu}
\affiliation{Stanford Synchrotron Radiation Lightsource, SLAC National Accelerator Laboratory, Menlo Park, California 94025, USA}

\date{\today}

\begin{abstract} 
The presence of a small concentration of in-plane Fe dopants in La$_{1.87}$Sr$_{0.13}$Cu$_{0.99}$Fe$_{0.01}$O$_4$ is known to enhance stripe-like spin and charge density wave (SDW and CDW) order, and suppress the superconducting $T_c$. Here, we show that it also induces highly two-dimensional (2D) superconducting correlations that have been  argued to be signatures of a new form of superconducting order, so-called pair-density-wave (PDW) order. In addition, using the resonant soft x-ray scattering, we find that the 2D superconducting fluctuation is strongly associated with the CDW stripe. In particular, the PDW signature first appears when the correlation length of the CDW stripe grows over eight times the lattice unit ($\sim$ 8$a$). These results provide critical conditions for the formation of PDW order.
\end{abstract}


\maketitle


It has been more than 30 years since high-temperature superconductivity (HTSC) was discovered in 1986 \cite{Muller}. Although tremendous progress has been made in understanding the fundamental mechanism of HTSC, especially in the high-$T_{\rm c}$ cuprates, the complexity of their phase diagrams \cite{Fradkin} has complicated the problem, hence there still remain many open issues. Some of the remaining profound questions are: Why is the superconducting transition temperature ($T_{\rm c}$) so high? How are different electronic and magnetic orders intertwined with HTSC? What is the pseudogap and how is it related to HTSC? 
Another interesting concerns is on the existence and characteristics of a possible new form of pairing, a so-called pair density wave (PDW), that potentially intertwines with several of the complex orders observed in the high-$T_{\rm c}$ cuprates \cite{Fradkin2}. Indeed, it has even been suggested that a fluctuating version of a PDW may be responsible for the famous pseudogap phenomenology \cite{Keimer,Lee,Zelli,Yu,Norman}.

In 2007, Li $et~al$. \cite{Li} reported the existence of two-dimensional (2D) superconductivity in a high-$T_{\rm c}$ cuprate, La$_{1.875}$Ba$_{0.125}$CuO$_4$ (LBCO). This, and closely related but less clear-cut results by Tajima $et~al$. \cite{Tajima}, have been identified as signatures of PDW order \cite{Himeda,Berg}. The most dramatic evidence of this unusual superconducting order is the existence of a range of temperatures, $T_c$ $\approx$ 4 K $<$ $T$ $<$ $T_{2}$ $\approx$ 17 K, in which the system exhibits an apparently anisotropic superconducting state. Specifically, in this temperature range, despite the existence of sufficiently strong 2D superconducting correlations that the resistivity with the current flowing in the CuO$_2$ planes  is immeasurably small, interlayer Josephson coupling is apparently absent, i.e., the $c$-axis resistivity remains large. Here $T_c$ is the transition temperature to the true superconducting (Meissner) state. Furthermore, in a somewhat higher range of temperatures $T_{2}$ $<$ $T$ $<$ $T_{1}$, a fluctuating version of the same state, with in-plane resistivity one or more order of magnitude smaller than the normal state resistivity, onsets below a temperature, $T_{1}$ which is approximately equal to $T_{\rm sdw}$ $\approx$ 40 K \cite{John,Hucker}, the onset temperature of  spin density wave (SDW) stripe order in the CuO$_2$ planes. Since the charge density wave (CDW) stripe develops at a still higher temperature $\approx$ 54 K \cite{Abbamonte,Homes,YKim,Miao} and thus coexists with the SDW at 40 K \cite{Hucker}, it was suggested that this PDW signature originates from an intertwining of the HTSC, SDW and CDW orders \cite{Fradkin2}. However, although the correlation between the SDW and CDW has been observed in multiple x-ray scattering measurements \cite{Hucker,Abbamonte,Homes,YKim,Miao}, no direct experimental evidence has unambiguously supported the association of this with PDW order.

More recently, scanning tunneling microscopy (STM) experiments in the halo region surrounding the vortex cores of Bi$_2$Sr$_2$CaCu$_2$O$_{8+\delta}$ (Bi2212) was performed by Edkins $et~al.$ \cite{Edkins}, aimed to detect the field induced PDW. Through enhanced real- and $Q$-space resolutions in the STM, they revealed a CDW with a period of approximately eight times the lattice unit ($\sim$ 8$a$), twice that of the normal CDW period ($\sim$ 4$a$). This observation is also thought to be a signature of a PDW. More detailed comparisons between microscopic theory and STM results have provided additional support for the existence of PDW order in Bi2212 \cite{Wang,Dai,Choubey}. However, the observation of PDW order through STM in Bi2212 is distinct from that in the LBCO system, if for no other reason than that there is presently no evidence of any role of SDW in the Bi2212 case. Although (rather short-range correlated) CDW order coexists with the superconducting phase of Bi2212 \cite{Neto,Chaix}, it doesn't have a phase overlapping with the magnetic order in the phase diagram \cite{Sterpetti,Drozdov}.

In this Letter, we explore new evidence of the putative PDW signature and its relation with other orders, such as CDW and SDW, in a La-based high-$T_{\rm c}$ cuprate La$_{1.87}$Sr$_{0.13}$Cu$_{0.99}$Fe$_{0.01}$O$_4$ (LSCFO) \cite{Supple}. It is important to stress that the crystal structure \cite{Achkar} of this material (i.e., the low temperature orthorhombic  - LTO - structure) is different from that of LBCO (i.e., the low temperature tetragonal - LTT). The LTT structure is known to suppress $T_c$ and enhance SDW and CDW order \cite{Abbamonte,Homes,YKim,Miao,Achkar}. In the present case, the CDW and SDW correlations are instead ``pinned'' and $T_{\rm c}$ is reduced by Fe doping rather than by the LTT structure. The pinned behavior is demonstrated by both the integrated intensity and peak position of CDW with the development of SDW. For this study, we carried out both transport measurements and resonant soft x-ray scattering (RSXS) around the Cu $L_3$-edge. Through the resistivity measurements along both the crystalline $ab$-axis (i.e., CuO$_2$ plane) and the $c$-axis and the comparison between them, we have identified a state of 2D superconducting fluctuations  that develops below $T_1$ $\approx$ 32 K. The RSXS measurement revealed that CDW short-range order (CDW-SRO) transforms into the CDW stripes with the development of SDW stripes below $T_{\rm sdw}\approx 50$ K. In contrast with the case in LBCO, $T_{\rm sdw}$ is significantly larger than $T_1$, although the onset of SDW order is visibly rounded, and becomes notably sharper around $T_1$. We further observed that the correlation length of the CDW stripe reaches around 8$a$, as the normal state undergoes a crossover to the 2D fluctuating state. While we do not observe any apparent vanishing of the in-plane resistivity at a temperature $T_1$ $>$ $T_{\rm c}$ $\sim$ 5.7 K \cite{Suzuki,Fujita}, we do see a form of crossover (discussed below) that is in some ways analogous to $T_1$ in LBCO. In short, the similarities with the observed relations between the different ordering phenomena in the present system strongly corroborate the intertwined character of the different orders, and correspondingly the association of it with PDW formation.  


A sizable LSCFO single crystal was grown by the standard floating-solvent traveling-zone method. The grown crystal was annealed in 1 bar of O$_2$ gas to minimize oxygen deficiencies. The doped Fe ions are in Fe$^{3+}$ state (3$d^5$, \textbf{S} = 5/2), which have higher valence state than the Cu atoms by one, resulting in a reduction of one hole per doped ion. Thus, the concentration of hole (i.e., doping, $p$) in our LSCFO is 0.12. We confirm that this crystal does not undergo the LTT by temperature dependence of the (0 1 0) Bragg peak \cite{Fujita}.

\begin{figure}[t]
\begin{center}
\includegraphics[width=0.45\textwidth]{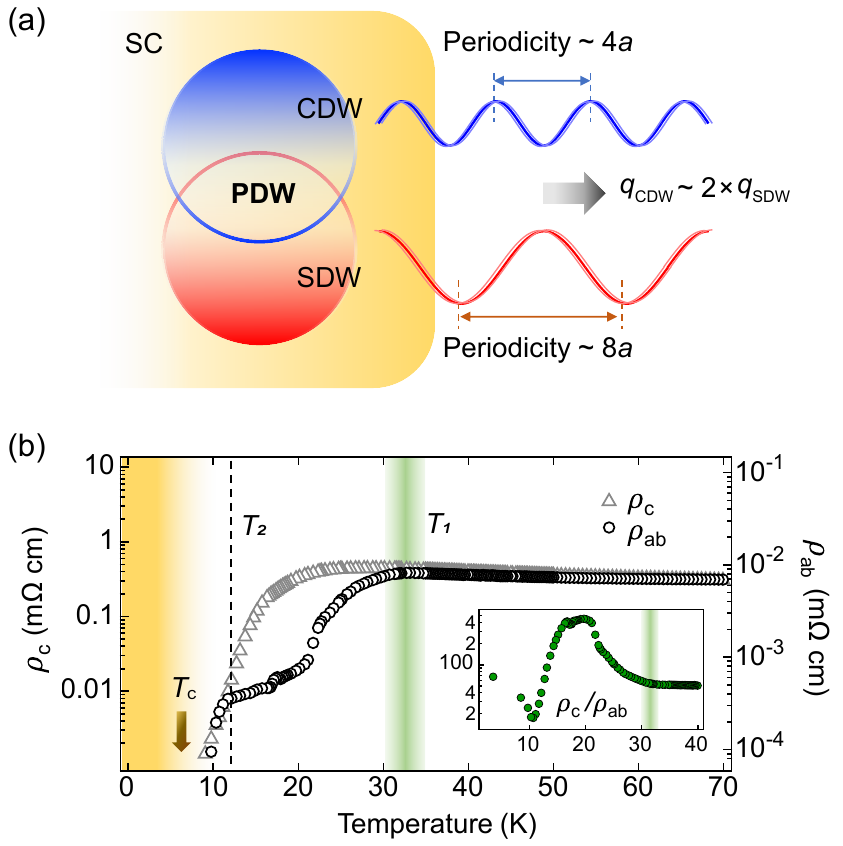}
\caption{(color online) (a) A schematic diagram of CDW and SDW stripes, and their intertwining that results in PDW order, implying the importance of their respective modulation phases. (b) Temperature dependence of the zero-field resistivity. $\rho_{\rm c}$ is along the $c$-axis and $\rho_{\rm ab}$ is parallel to the CuO$_2$ planes. The green shade and dashed line denote the first and second transitions in $\rho_{\rm ab}$, respectively, above the Meissner state of both $\rho_{\rm ab}$ and $\rho_{\rm c}$ ($T_{\rm c}$ $\sim$ 5.7 K). The inset shows the ratio of $\rho_{\rm c}$ to $\rho_{\rm ab}$ as a function of temperature.} \label{Fig1}
\end{center}
\end{figure}

We first measured the resistivity ($\rho$) of our sample aiming to explore a PDW signature in LSCFO \cite{Supple, PPMS}. This measurement was motivated by the fact that PDW order was observed in LBCO \cite{Li} and that La-based cuprates share the same mutual relationship between the CDW and SDW orders \cite{Hucker,Miao,Yamada,Fujita,Wen}. As illustrated in Fig.\ \ref{Fig1}(a), it is well-known that the ordering $q$-vectors of the CDW and SDW ($q_{\rm cdw}$ and $q_{\rm sdw}$) are related by $q_{\rm cdw}$ $\sim$ 2$\times q_{\rm sdw}$ \cite{Wen,Yamada}. It was demonstrated that the LSCFO system also follows this relation \cite{Fujita}. Figure 1(b) shows the resistivity along the $c$-axis ($\rho_{\rm c}$) as well as that  parallel to the CuO$_2$ plane ($\rho_{\rm ab}$), which reveals an anisotropic behavior occurring around 32 K ($T_1$). Below the $T_1$, the value of $\rho_{\rm ab}$ gets as low as $\sim$ 10$^{-3}$ m$\Omega$ cm, while $\rho_{\rm c}$ $\sim$ 10$^0$ m$\Omega~\rm{cm}$ slightly increases. The anisotropy is even more obvious in the $\rho_{\rm c}$/$\rho_{\rm ab}$ ratio plot (see the inset), showing an increase of one order of magnitude. This result shows strong resemblance to the resistivity behavior in LBCO \cite{Li} and Zn-doped LBCO \cite{Lozano}, lending an interpretation that our LSCFO sample exhibits 2D fluctuating superconductivity (i.e., PDW fluctuation) around $T_1$. Note that we also observed magnetic field dependent $\rho_{\rm ab}$ \cite{Supple}. A second drop of $\rho_{\rm ab}$ occurs around 11 K ($T_2$), deducing that the PDW fluctuation may change to a pure 2D superconducting state. Note that $T_2$ is still higher than the bulk $T_{\rm c}$ $\sim$ 5.7 K of LSCFO \cite{Supple}.

\begin{figure*}
\begin{center}
\includegraphics[width=0.96\textwidth]{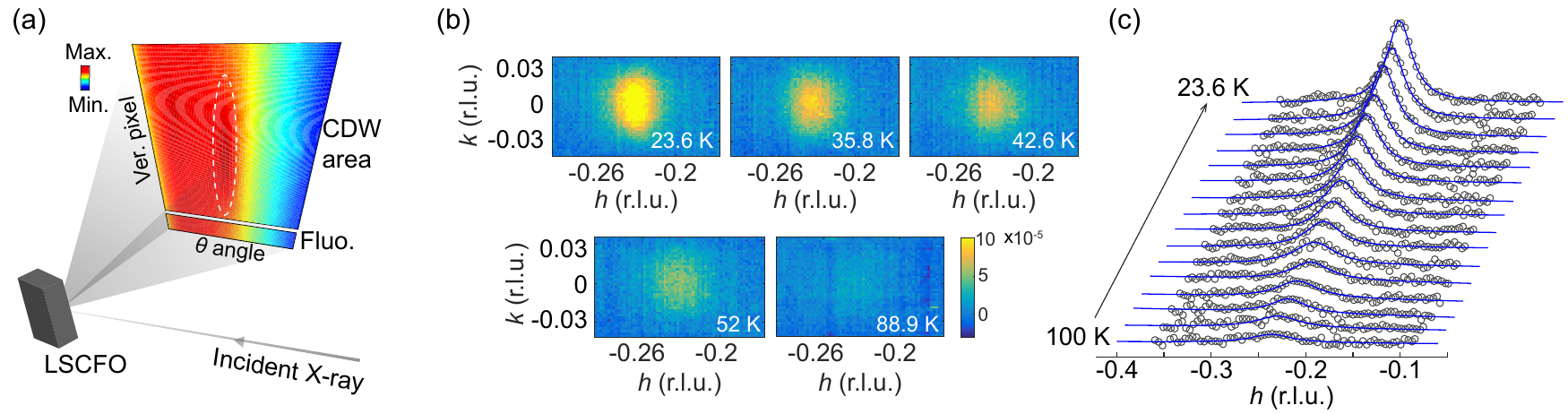}
\caption{(color online) (a) A schematic sketch showing the RSXS experiment on LSCFO using an area detector (CCD). At the fixed CCD angle (150$^\circ$), $\theta$-scans were performed. The CDW signal is aligned on the center of the CCD detector. The bottom edge area of the image is regarded as the fluorescence (Fluo.) background-dominated region. (b) RSXS patterns in $h$/$k$ space of LSCFO for various sample temperatures after subtracting the fluorescence background. The CDW patterns are centered at $q_{\rm cdw}$ $\sim$ (-0.233, 0, $l$). Below $\sim$ 50 K, the pattern tends to elongate along the $k$-direction. (c) Projected scattering profiles along the $h$-direction as a function of temperature. The solid lines are Lorentzian fits. 
} \label{Fig2}
\end{center}
\end{figure*}

\begin{figure}[b]
\begin{center}
\includegraphics[width=0.46\textwidth]{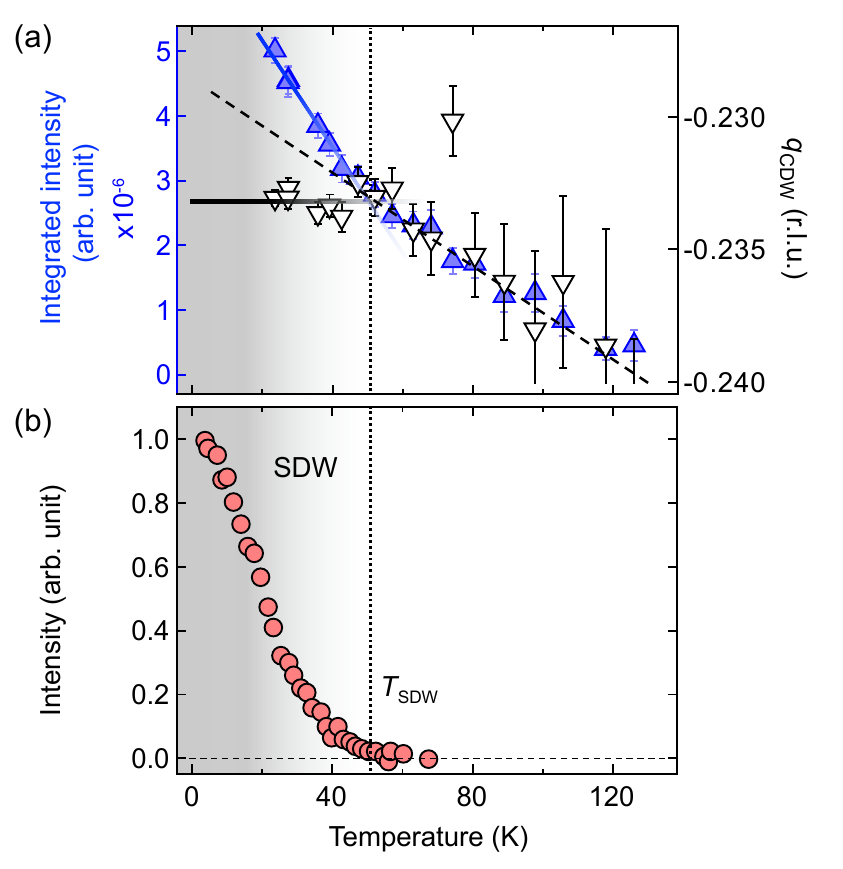}
\caption{(color online) (a) Integrated intensity (up-triangles) of the CDW order (left y-axis) and its wavevector $q_{\rm cdw}$ (down-triangles, right y-axis) as a function of temperature. Below $\sim$ 50 K (denoted by the vertical, dotted line), both the intensity and the $q_{\rm cdw}$ incommensurability show a different trend. The error bars represent 1 standard deviation (s.d.) of the fit parameters. (b) Temperature dependence of the SDW peak intensity measured by elastic neutron scattering. This data was taken from Ref.\cite{Fujita}. The vertical, dotted line at $\sim$ 50 K denotes the onset of SDW order.
} \label{Fig3}
\end{center}
\end{figure}

Based on the observation of PDW fluctuation in LSCFO, as a next step we investigated its relationship with CDW and SDW orders in our sample. We measured the SDW order using the elastic neutron scattering as reported in Ref.\cite{Fujita}. For CDW measurements, we employed resonant soft x-ray scattering (RSXS) approach that has demonstrated its capability to detect small changes in weak CDW signals \cite{Abbamonte, Wu, Thampy, Wen}.

Figure\ \ref{Fig2}(a) shows a schematic configuration for the RSXS measurements, where scattered x-ray signals were measured as a function of sample-angle ($\theta$) by a CCD detector. During the $\theta$-scan, the detector image covers scattering intensities from the well-aligned scattering plane ($h$, 0, $l$) near the detector center as well as from off-scattering planes ($h$, $\pm k$, $l$) at the top/bottom area of the CCD. As depicted in the figure, we used the bottom part of the CCD images to subtract out the fluorescence background signal. After the background subtraction, we achieved clear CDW maps along the $h$-/$k$-direction centered at $q_{\rm cdw}$ $\sim$ (-0.23, 0, $l$) reciprocal lattice units (r.l.u.) (see Fig.\ \ref{Fig2}(b) for the background-subtracted scattering maps at selected temperatures). As shown in the figure, the CDW peaks are elongated along the $k$-direction as the temperature is decreased below $T_{\rm sdw}$ $\sim$ 50 K. This observation is consistent with previous reports on LSCO where CDW signals showed peak-splitting through intertwining between striped CDW and SDW \cite{Wen,Kimura,Thampy}, suggesting that CDW stripes are also formed in LSCFO. Figure\ \ref{Fig2}(c) shows projected CDW signals along the $h$-direction as a function of temperature. We fitted the projected CDW peaks with the Lorentzian function (blue lines in this figure) and the results are summarized in Fig.\ \ref{Fig3}.

Similarly to the LSCO case \cite{Wen}, CDW order in LSCFO continuously develops as the sample is cooled down from above 120 K. As shown in Fig.\ \ref{Fig3}(a), the integrated intensity (left $y$-axis) of the CDW peak keeps growing with decreasing temperature. Notably, the growing trend shows a change around 50 K, indicating a transition of the CDW. Note that the transition behavior is not sharp because the characteristics of density wave orders is typically glassy in the cuprates. The high temperature CDW-SRO phase is transformed into the CDW stripe around this temperature, as was the case for LSCO \cite{Wen}. Furthermore, this slope change in the integrated intensity coincides with a slope change in the wavevector of CDW order ($q_{\rm cdw}$, right y-axis). The $q_{\rm cdw}$ decreases with decreasing temperature, but is locked in below 50 K. Considering the LBCO case \cite{Miao}, this locking in LSCFO can be explained by the development of the SDW stripe. As shown in Fig.\ \ref{Fig3}(b), 50 K is the temperature where the SDW stripe develops \cite{Fujita}. These results indicate that the CDW stripe phase is developed by intertwining with the SDW. However, the discrepancy between this intertwining temperature and the onset temperature of PDW fluctuation obtained by the transport measurement (Fig.\ \ref{Fig1}(b)) calls for explanation.

\begin{figure}[t]
\begin{center}
\includegraphics[width=0.46\textwidth]{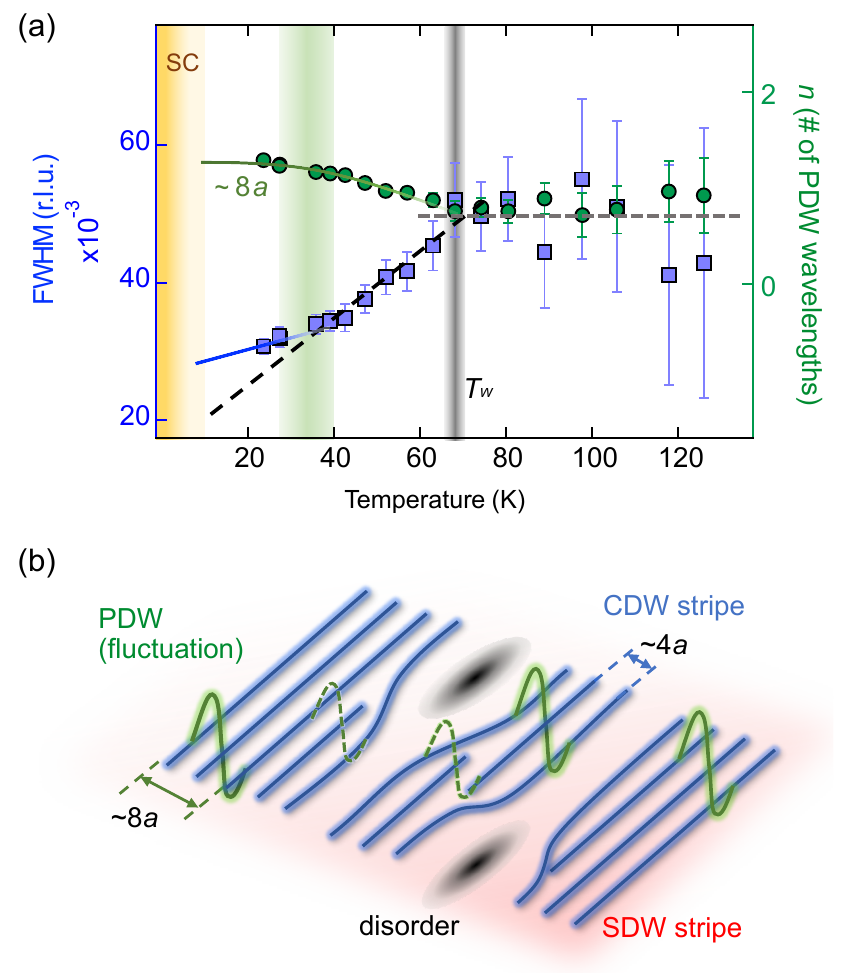}
\caption{(color online) (a) FWHM (squares) of the CDW order (left $y$-axis) and the number of wavelengths (circles) for the PDW order (right $y$-axis). The number of wavelengths ($n$) is estimated by $\xi_{\rm CDW}$/$8a$, where $\xi_{\rm CDW}$ is calculated as 2/FWHM. Note that the FWHM and $n$ show the same trend because they are inversely proportional with each other; they are plotted with the different axis scale. The vertical grey shade denotes a characteristic temperature ($T_{\rm w}$), at which the FWHM of CDW starts to change while cooling down. The vertical green shade denotes another slope change in FWHM, close to the first anisotropic transition ($T_1$) in $\rho_{\rm ab}$. Below $T_1$ $\sim$ 32 K, $n$ is larger than 1. (b) A schematic diagram of a CuO$_2$ plane in LSCFO illustrating the formation of the PDW order. The PDW (fluctuation) starts to form when the domain size of CDW is over $\sim$ 8$a$. The dashed green curves represent a situation where PDW cannot be formed because of insufficient $\xi_{\rm CDW}$. The black shades denote intrinsic disorders.
} \label{Fig4}
\end{center}
\end{figure}

To understand the discrepancy, we further scrutinize the CDW correlation lengths ($\xi_{\rm CDW}$) in LSCFO. Figure\ \ref{Fig4}(a) shows the full width at half maximum (FWHM) of the CDW peak as a function of temperature (left $y$-axis). For $T$ $>$ 70 K, the FWHM does not change within the error, representing a $T$-independent region. This $T$-independent FWHM corresponds to $\xi_{\rm CDW}$ $\sim$ 23(2) \AA, which is longer than the CDW wavelength, $\xi_{\rm CDW}$ $\sim$ 4$a$ $\sim$ 15 \AA. For $T$ $<$ 70 K, on the other hand, the FWHM starts to decrease ($\xi_{\rm CDW}$ increases) with decreasing temperature. Following the discussion in Ref.\cite{Wen}, we assign 70 K to the characteristic temperature ($T_{\rm w}$).

The FWHM behavior shows another noteworthy feature. The increasing trend of $\xi_{\rm CDW}$ below the $T_{\rm w}$ appears to undergo another slope change around 34$\pm$3 K \cite{Supple}, which is close to the crossover temperature ($T_1$) of PDW fluctuating states. A very interesting observation is that the measured $\xi_{\rm CDW}$ at this temperature is around 35(2) \AA, which is close to the PDW wavelength, $\lambda_{\rm PDW}$ $\sim$ 8$a$ $\sim$ 30 \AA. For easier comparison between the two wavelengths, we plotted the ratio $n$ = $\xi_{\rm CDW}$/$\lambda_{\rm PDW}$ on the right $y$-axis in Fig.\ \ref{Fig4}(a). At $T_1$, $n$ is slightly over one. This observation implies that $\xi_{\rm CDW}$ $\geq$ $\lambda_{\rm PDW}$ is a critical condition, together with the presence of the CDW stripe, for the formation of PDW order, as highlighted in Fig.\ \ref{Fig4}(b). When the domain size of CDW stripe is smaller than 8$a$, PDW fluctuation cannot be formed (green dashed line in Fig.\ \ref{Fig4}(b)) even when both the SDW and CDW are present. These conditions are fully compatible with the LBCO case, where PDW was readily formed when CDW stripes were formed around the SDW temperature \cite{Li}, because at this temperature the $\xi_{\rm CDW}$ is fairly long and already meets the $\xi_{\rm CDW}$ $\geq$ $\lambda_{\rm PDW}$ condition due to the LTT \cite{Hucker,YKim}.


We now discuss the implications of our findings. First, the new evidence of the 2D superconducting signature in this LSCFO system means the PDW order observed in LBCO is not a consequence of some material science specific to LBCO, but could be ubiquitous in La-based cuprates. Second, our findings, especially the critical prerequisite conditions for PDW, are compatible with the two PDW (or its fluctuation) signatures observed in LBCO and Bi2212 \cite{Li,Edkins,Du}, which are seemingly contradicting each other. Considering the enhancement of $\xi_{\rm CDW}$ below $T_{\rm c}$ under the magnetic field as a result of the well-known competition between the CDW and SC orders \cite{Ghiringhelli,Chang,Gerber}, the vortex halo region in Bi2212 under the magnetic field is expected to have a large CDW domain size, satisfying the correlation length condition to trigger the PDW. In La-based cuprates, this condition is met by a different mechanism; the SDW stripe order helps develop a long-ranged CDW stripe through intertwining. The different situations between Bi2212 and La-based cuprates lead to another critical implication that of the two orders, namely the CDW and SDW stripes, the CDW stripe (or $\xi_{\rm CDW}$) is the more essential ingredient for prompting the PDW. This implication is also supported by the observation that for LSCO with 10\% hole-doping, the measured $c$-axis superconducting coherence is abruptly quenched with a magnetic field (smaller than $H_{\rm c2}$) while the same magnetic field strongly enhances the stripe order \cite{Schafgans,Lake}. Finally, one may extend the implications to a broader context of the pseudogap phenomenology in the high-$T_{\rm c}$ cuprates \cite{Keimer,Lee,Zelli,Yu,Norman}. There are some pieces of evidence on the unusual gap features that suggest a PDW fluctuation may be responsible for the pseudogap phenomenology \cite{Lee,Zelli,Yu,Norman}. Nevertheless, due to different energy and temperature scales between these two phenomena, it is still early to connect them directly. On the other hand, we may contemplate that the CDW could play a bridging role between them, because the CDW is associated with each phenomenon. For example, in the LSCO case, the upper critical doping boundary of the CDW phase is strikingly close to the pseudogap critical doping \cite{Wen,Badoux,Taillefer}, supporting the close relationship between the CDW and the pseudogap.

In summary, we carried out resistivity and Cu $L$-edge RSXS measurements on a single crystal of La-based cuprate, La$_{1.87}$Sr$_{0.13}$Cu$_{0.99}$Fe$_{0.01}$O$_4$. We found new evidence of PDW in this system. We also found that its emergence is attributed primarily to the CDW stripe rather than the SDW stripe, although the CDW stripe is developed through intertwining with SDW stripes in this system. In particular, the PDW signature starts to appear when the correlation length of the CDW stripe grows over 8$a$. These results indicate the critical role of the CDW stripe for the formation of PDW order in high-$T_{\rm c}$ cuprates. Moreover, our findings support the possibility that the PDW is ubiquitous in every high-$T_{\rm c}$ cuprate system.

We thank Steven A. Kivelson, Hoyoung Jang, and Jiajia Wen for valuable discussions and comments. Resonant soft x-ray experiments were carried out at the SSRL (beamline 13-3), SLAC National Accelerator Laboratory. This study at the SSRL/SLAC is supported by the U.S. Department of Energy, Office of Science, Office of Basic Energy Sciences under contract no. DE-AC02-76SF00515. The transport (resistivity) measurements were performed at IMR/Tohoku University funded by a Grant-in-Aid for Scientific Research (A) (grant no. 16H02125) and Scientific Research (C) (grant no. 16K05460).


\begin{thebibliography}{}
\bibitem{PPMS} About the transport measurement -- Electrical resistivity was measured down to 2.5 K using a standard dc-four-terminal-probe method with a physical property measurement system (PPMS: Quantum Design).
\end{thebibliography}
\end{document}